# PON-based connectivity for fog computing

Abdullah M. Alqahtani, Sanaa H. Mohamed, Taisir E. H. El-Gorashi and Jaafar M. H. Elmirghani
School of Electronic & Electrical Engineering, University of Leeds, LS2 9JT, United Kingdom

**ABSTRACT**
Fog computing plays a crucial role in satisfying the requirements of delay-sensitive applications such as connected vehicles, smart grids, and actuator networks by moving data processing close to end users. Passive optical networks (PONs) are widely used in access networks to reduce the power consumption while providing high bandwidth to end users under flexible designs. Typically, distributed fog computing units in access networks have limited processing and storage capacities that can be under or over utilized depending on instantaneous demands. To extend the available capacity in access network, this paper proposes a fog computing architecture based on SDN-enabled PONs to achieve full connectivity among distributed fog computing servers. The power consumption results show that this architecture can achieve up to about 80% power savings in comparison to legacy fog computing based on spine and leaf data centers with the same number of servers.

**Keywords**: Fog computing, Passive optical networks (PON), Software-defined networking (SDN), Fog computing, Energy efficiency, Arrayed waveguide grating routers (AWGRs), Mixed integer linear programming (MILP).

## 1. INTRODUCTION

Time-sensitive applications that require quick response for data processing, such as some internet of things (IoT) applications, connected vehicle, smart grid and actuators networks pose challenges to computing systems including delay and power consumption challenges. [1]–[5]. Previously, cloud computing, placed in the core network, was used to meet the needs of individuals and organizations by offering a range of computing services, data storage and network management functions. However, to support time-sensitive applications, the computing systems should be placed close to the users to provide the required quick response. Currently, fog computing, which is placed in the access network, is considered to provide lower latency and lower networking power consumption compared to cloud computing [6], [7]. Moreover, passive optical networks have been deployed in access networks and considered also in data centers to reduce the power consumption and to provide high bandwidth [8]–[14].The authors in [15]–[17] have proposed the use of PONs to facilitate the inter and intra racks communication in data centers to reduce the power consumption. Software defined networking (SDN) concepts can be used in PONs to provide flexibility in wavelength routing and assignment. It was also considered for PON-based data centers to facilitate the inter-rack communications [18]–[20].

Minimizing the power consumption is one of the current challenges in networking and computing systems [21]–[23]. The authors in [24]–[28] have optimized the network and computing energy efficiency using virtualization techniques in cloud computing systems. The authors in [29]–[35] tackled the design of networking architectures to minimize the power consumption. The authors in [36]–[41] have focused on optimizing the distribution of content in multiple locations to achieve energy efficiency in networking and computing systems. Also, the impact of big data has been tackled in [42]–[46] to identify an optimal processing system. Moreover, the author in [47] addressed the reduction in brown power consumption by utilizing renewable energy sources in networking and computing systems. In this paper, we propse PON technology to connect multiple Fog Computing units and determine the resultant power consumption reduction. We extended one of the PON-based data center designs proposed in [15] to include multiple PON cells, each cell represents a fog computing unit as shown in Fig 1. The proposed architecture represents an extendible PON-based data center in the access network where a number of fog computing units have full PON-based connectivity, which allows storage and processing capabilities to be borrowed and shared among the PON processing cells. At the same time, we aim to reduce the number of intermediate hops between different cells to achieve low-latency communication.

The reminder of this paper is organized as follows. Section 2 explains the proposed architecture used to achieve full connectivity between distributed fog computing units along with its wavelength routing and assignment. Section 3 presents a power consumption study that compares the power consumption of the proposed architecture with the power consumption of a spine and leaf data center architecture. Finally, Section 4 concludes the paper and outlines avenues for future work.

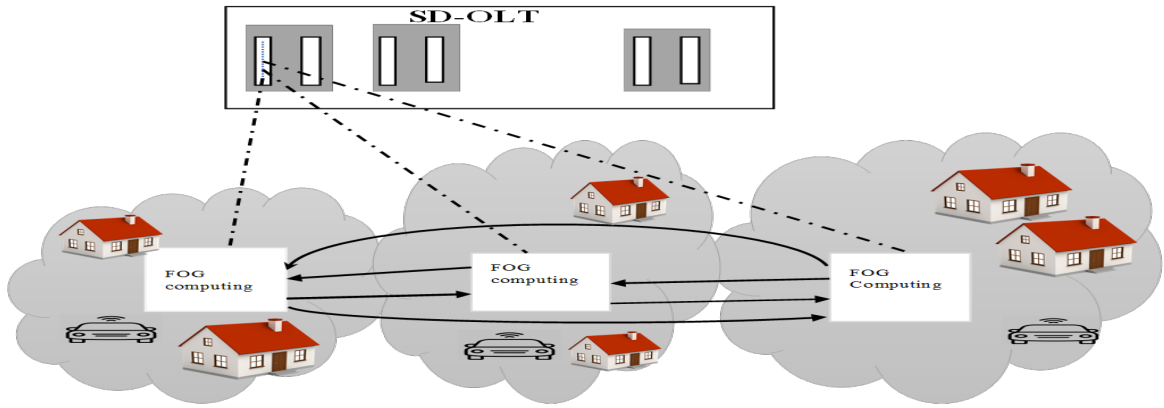

**Figure 1 Fog Computing based on SDN-enabled PON technology**

## 2. Fog Computing based on SDN-enabled WDM-PON technology

Figure 2 shows the proposed fog computing architecture. A group of racks with servers (i.e. a fog computing unit) is considered a PON cell and the architecture fully interconnects three PON cells. All cells are connected to one SDN-enabled Optical Line Terminal (OLT). In Figure 2, each fog computing unit is composed of two racks, each represents a PON group and hosts 16 servers since typically the space is small for a fog computing unit [48]. Each server is connected to a wavelength tunable Optical Network Unit (ONU). Communication between servers located in the same rack (i.e. intra-rack communication) is achieved passively as in [13] via fibre Bragg grating or a star reflector, while communication between the servers located in different racks depicted as inter-rack communication is achieved via an OLT switch or Arrayed Waveguide Grating Routers (AWGRs). We assume that the distance between the SDN-enabled OLT and the PON cells do not exceed 20 km, as the typical distance in PON access networks [49]. An OLT chassis can typically host up to 18 cards, two of which are reserved for the switching matrix and control. Each card can have up to 16 GPON ports, with split ratios of up to 128 ONUs per port. This means that one port can serve up to 128 servers and one card can serve up to 2048 servers [50]. To facilitate the inter-PON-Cell-communication, as shown in Figure 2, each AWGR should have connections to all AWGRs placed in other PON cells. The connections between AWGRs in Figure 2 are arranged to achieve the full connectivity. In addition, each AWGR should have a connection with the SDN-enabled OLT (SD-OLT) through Arrayed Waveguide Grating (AWGs) multiplexer to send requests and receive instruction to communicate with other PON groups. The SDN-enabled OLT is connected with an SDN controller to receive instructions related to routing and wavelength assignment.

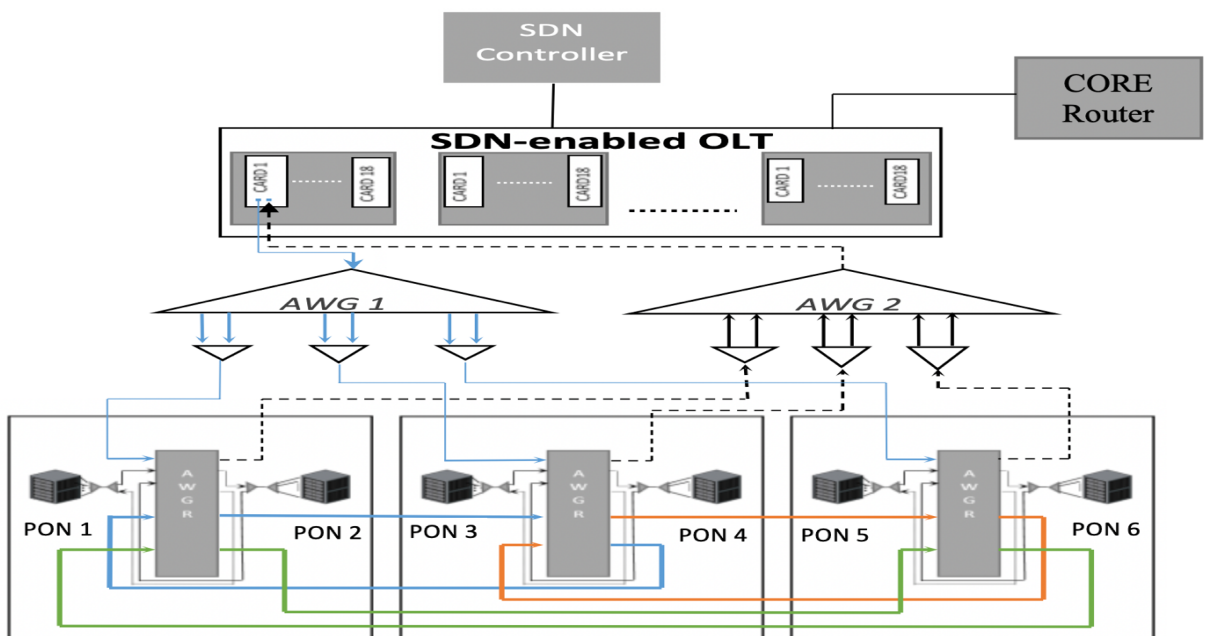

**Figure 2 PON Cellular connectivity based on AWGR**

## 2.1 Result and Discussion

For the proposed connections among the AWGs and AWGRs presented in Figure 2, a Mixed Integer Linear Programming (MILP) model was used to optimize the wavelength assignment and routing for the communication between the PON groups and with the SDN-enabled OLT. Table1 shows the results of the wavelength assignments and routing for inter-communication between PON groups and with OLT.

**Table 1: Routing map for inter-PON group communications**

|  |  | GROUP 1 | GROUP 2 | GROUP 3 | GROUP 4 | GROUP 5 | GROUP 6 | OLT |
|---|---|---|---|---|---|---|---|---|
| PON CELL 1 | GROUP 1 | - | $\lambda_3$ | $\lambda_2$ | $\lambda_5$ | $\lambda_4$ | $\lambda_1$ | $\lambda_6$ |
|  | GROUP 2 | $\lambda_3$ | - | $\lambda_1$ | $\lambda_4$ | $\lambda_2$ | $\lambda_6$ | $\lambda_5$ |
| PON CELL 2 | GROUP 3 | $\lambda_2$ | $\lambda_5$ | - | $\lambda_1$ | $\lambda_6$ | $\lambda_3$ | $\lambda_4$ |
|  | GROUP 4 | $\lambda_4$ | $\lambda_1$ | $\lambda_6$ | - | $\lambda_5$ | $\lambda_2$ | $\lambda_3$ |
| PON CELL 3 | GROUP 5 | $\lambda_1$ | $\lambda_6$ | $\lambda_5$ | $\lambda_3$ | - | $\lambda_4$ | $\lambda_2$ |
|  | GROUP 6 | $\lambda_5$ | $\lambda_2$ | $\lambda_4$ | $\lambda_6$ | $\lambda_3$ | - | $\lambda_1$ |
|  | OLT | $\lambda_6$ | $\lambda_4$ | $\lambda_3$ | $\lambda_2$ | $\lambda_1$ | $\lambda_5$ | - |

In the design shown in Figure 2, which includes 6 PON groups, 6 wavelengths are needed to achieve full connectivity. The servers located within one PON cell but in different groups or within different cells are able to communicate with each other via AWGRs to achieve inter-rack communication in one extendible PON cellular architecture. For example, if a server in PON group 3 needs to communicate with a server in PON group 5, both of them placed in different PON cells, the communication proceeds as follows: First, the server in PON group 3 sends a request message to the OLT using $\lambda_4$. When the SDN-enabled OLT authorizes the communication, it sends a control message containing the wavelength assignment to the servers in PON group 3 and PON group 5 so that they can communicate with each other, using $\lambda_3$ and $\lambda_1$ respectively. Based on the wavelength assignment sent from the SDN-enabled OLT, the two servers are able to communicate with each other by tuning their transceivers to $\lambda_6$.

## 3. The Power Consumption of the proposed Fog Connectivity Based on SDN-enabled OLT

This section discusses the results of a power consumption study which compares fog computing connectivity based on SDN-enabled OLT architecture with a spine and leaf data center architecture implemented in the access layer also. The spine and leaf architecture, shown in Figure 4, is composed of two layers: the leaf layer, which connects all the servers and storage, and the spine layer, which is the core layer of the architecture and is responsible for interconnecting all access (leaf) switches [51].

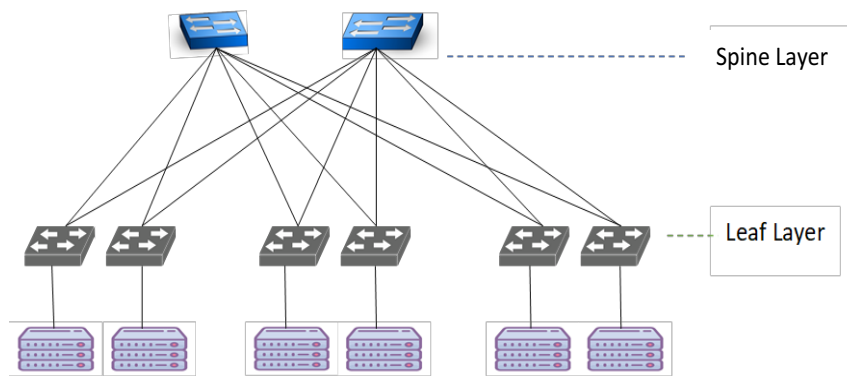

**Figure 3: Spine and Leaf Architecture**

The power consumption of the equipment used in the proposed architecture and spine and leaf architecture are shown in Table 2.

Table 2: Power Consumption of Different Equipment.

| Equipment | Power consumption |
|---|---|
| OLT, GPON line card [50] | 90W |
| Tuneable ONU [52] | 2.5W |
| Spine Switch [53] | 660W |
| Leaf Switch [54] | 508W |
| Server Transceiver [55] | 3W |

The proposed design reduces power consumption by 80 percent when 96 servers are considered compared to the spine-and-leaf architecture. The underlying reason for the high-power consumption of spine and leaf architecture is the large number of switches involved in such interconnections. In the proposed architecture, these switches are replaced by passive devices such as AWGRs, couplers and FBGs in order to reduce the power consumption. Using AWGRs to route traffic instead of SDN-enabled OLTs and commodity switches is a crucial aspect of reducing power consumption due to the passive nature of AWGRs. This approach also reduces the delay in inter-rack communication by avoiding the queuing in the OLT switch. The power consumption saved by the proposed architecture compared to spine and leaf is slightly reduced when the number of servers increases, as shown in Figure 5. This is because the power consumption of the switches used to build the spine-and-leaf layers does not increase linearly when the number of ports is increased. On the other hand, the power consumption of the proposed architecture increases linearly as the number of servers increases because the servers are equipped with tunable ONUs.

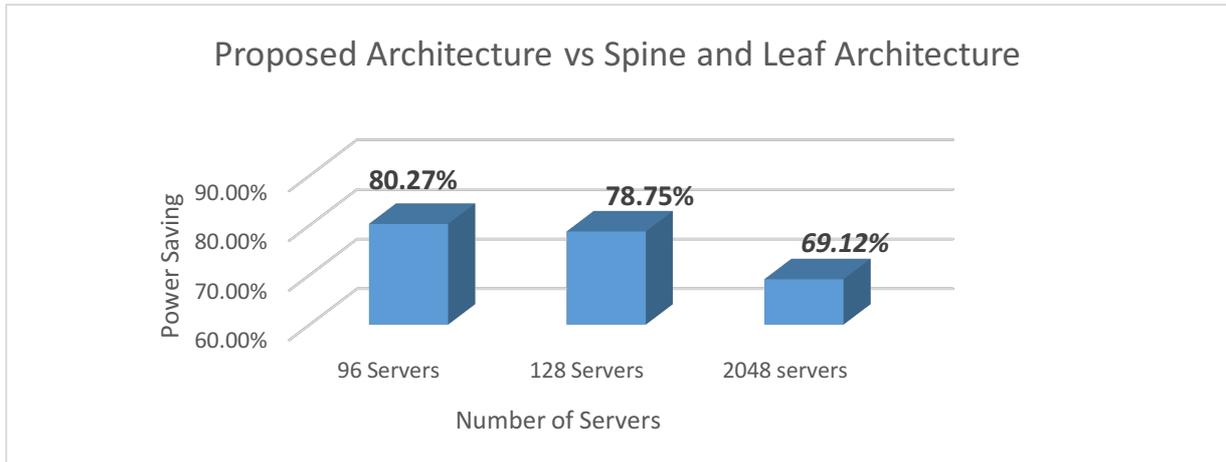

Figure 4 Power Savings for the proposed Fog arcitecture compared to Spine and Leaf arcitecture

## 4. CONCLUSIONS AND FUTURE WORK

In this paper, fog computing connectivity based on SDN-enabled OLT is proposed in order to support the server to server traffic in case processing has to be offloaded to neighbouring cells. The results obtained in the proposed architecture indicate that a power comsumption saving of up to about 80% is achieved in comparison to the spine-and-leaf architecture. Future work includes optimizing the resource allocation to minimize the power consumption in the proposed architecture and consideration of mobility-aware workload assignments for energy efficient fog computing based on PON architectures.

*Acknowledgments*

The authors would like to acknowledge funding from the Engineering and Physical Sciences Research Council (EPSRC) INTERNET (EP/H040536/1), STAR (EP/K016873/1) and TOWS (EP/S016570/1) projects. The first author would like to acknowledge the Government of Saudi Arabia and JAZAN University for funding his PhD scholarship. SHM would like to thank EPSRC for providing her Doctoral Training Award scholarship. All data are provided in full in the results section of this paper.